\documentclass[journal]{IEEEtran}
\usepackage{cite}
\usepackage{amsmath}
\usepackage{amssymb}
\usepackage{amsfonts}
\usepackage{graphicx}
\usepackage{float}
\usepackage{stfloats} 
\usepackage{subcaption}
\usepackage{multirow}  
\usepackage{array}     
\usepackage{graphicx}
\usepackage{epstopdf}
\usepackage{booktabs} 
\usepackage{caption} 
\ifCLASSINFOpdf
\else

\fi

\begin{document}

\title{Robust SCMA Codebook Design: A Hardware-Aware Autoencoder Approach}

\author{Zihao~Liu,~\IEEEmembership{Student~Member,~IEEE,}
        Zilong~Liu,~\IEEEmembership{Senior~Member,~IEEE,}
        Leila~Musavian,~\IEEEmembership{Senior~Member,~IEEE}
        
\thanks{The simulation code to reproduce the results in this paper is available at: https://github.com/ZihaoLiu258-source/Autoencoder-Based-OFDM-SCMA-Codebook-Design}        
\thanks{Zihao Liu, Zilong Liu and Leila Musavian are with the School of Computer Science and Electronic Engineering, University of Essex, Colchester CO4 3SQ, U.K. (e-mail: zl346810274@gmail.com; zilong.liu@essex.ac.uk; leila.musavian@essex.ac.uk).}% <-this % stops a space
}

\markboth{IEEE Wireless Communication Letters}%
{Liu \MakeLowercase{\textit{et al.}}: Robust SCMA Codebook Design: A Hardware-Aware Autoencoder Approach}

\maketitle

\begin{abstract}
Sparse code multiple access (SCMA) is a promising code-domain non-orthogonal multiple access scheme which is transmitted over orthogonal frequency division multiplexing (OFDM) to exploit multicarrier diversity. In practice, however, carrier frequency offset (CFO) and phase noise (PN) may disrupt the subcarrier orthogonality in OFDM-SCMA systems. Addressing this research problem from a new SCMA codebook design angle, we propose a hardware-aware end-to-end autoencoder that embeds differentiable CFO and Wiener PN layers into the training loop. Simulations show that the proposed codebook effectively 
suppresses the bit error floors caused by CFO and PN without requiring 
real-time phase tracking.
\end{abstract}


\IEEEpeerreviewmaketitle

\section{Introduction}

\IEEEPARstart{S}{PARSE} code multiple access (SCMA) building upon orthogonal frequency division multiplexing (OFDM) has emerged as a promising code-domain non-orthogonal multiple access scheme \cite{6666156, 9369968} for supporting massive machine-type communications (mMTC) in the next-generation wireless networks \cite{9693417}, offering high spectral efficiency and low latency \cite{6966170}.

The practical deployment of OFDM-SCMA is, however, often hindered by various hardware impairments. Among many others, carrier frequency offset (CFO) and oscillator related phase noise (PN) may destroy the subcarrier orthogonality \cite{A}, resulting in inter-carrier interference (ICI) and common phase error (CPE). Conventional codebook designs \cite{deka,li,zhang} are mostly designed by assuming zero CFO and zero PN. Consequently, under high CFO and PN, OFDM-SCMA may experience significant performance degradation and bit error floors due to phase rotations and ICI \cite{10342857}. Since deploying complex phase-tracking loops may be impractical for low-cost mMTC devices, such a physical layer vulnerability remains a critical bottleneck. While the codebooks in \cite{qu} offer certain resilience, OFDM transmission was not explicitly considered therein. Further, their designed codebooks lack robustness to practical impairments by assuming a simplified independent PN model rather than a time-varying Wiener process.

Recently, autoencoder (AE)-based deep learning (DL) has been successfully applied to physical layer optimization \cite{8054694} and SCMA codebook design \cite{zheng,luo}. Although these data-driven approaches often achieve good performance, most existing DL schemes rely on idealized channel models assuming perfect carrier synchronization and ideal oscillators. Due to lack of explicit modeling for radio frequency (RF) impairments during training, these codebooks remain susceptible to real-world phase mismatches.

To fill this research gap, we develop a hardware-aware end-to-end DL framework to design novel SCMA codebooks for combating CFO and PN in OFDM transmission. A differentiable Log-Message Passing Algorithm (MPA) is employed as the decoder to enable end-to-end gradient backpropagation. The main contributions are summarized as follows:

Firstly, we embed the CFO rotation and the time-varying Wiener PN as differentiable layers inside the autoencoder forward path. The key enabler is a reparameterization that recasts the stochastic PN random walk and the AWGN as deterministic functions of Gaussian variables, decoupling the randomness from the trainable codebook tensor. This converts the expected detection loss into a objective, thus allowing gradients carrying the impairment statistics to propagate back to the codebook.

Secondly, thanks to effective learning via a composite curriculum objective, we show that the autoencoder converges to a ring-based constellation that spreads codewords across distinct amplitude shells. Since CFO and PN manifest primarily as angular rotations, this radial separability acts as a phase buffer: codewords remain distinguishable under stochastic rotations, whereas conventional codebooks suffer symbol collisions and bit error floors. Crucially, this robustness is obtained through offline optimization with no additional receiver complexity.

\section{System Model}
Consider a downlink OFDM-SCMA system serving $J$ users over $K$ orthogonal subcarriers ($J > K$), with the overloading factor defined as $\lambda = J/K$.

\subsection{OFDM-SCMA Transmitter}
At the transmitter, a block of $\log_2(M)$ data bits from the $j$-th user ($j \in \{1, \dots, J\}$) is mapped to a $K$-dimensional complex sparse codeword $\mathbf{c}_j$ chosen from a codebook $\mathcal{C}_j \in \mathbb{C}^{K \times M}$, where $M$ is the constellation size. The codeword $\mathbf{c}_j$ has $d_j < K$ non-zero entries. The superimposed signal on the $k$-th subcarrier, denoted by $x_k$, is the sum of signals from all users:
\begin{equation}
x_k = \sum_{j=1}^{J} c_{j,k}, \quad k = 1, \dots, K,
\end{equation}
where $c_{j,k}$ is the element of codeword $\mathbf{c}_j$ on subcarrier $k$. To fully utilize the spectrum, the $N$ available subcarriers are divided into $Q = N/K$ independent SCMA transmission blocks without zero-padding. The $K$-dimensional frequency-domain symbol vectors from these $Q$ blocks are then concatenated to form the complete $N$-point frequency-domain vector $\mathbf{X}=[X_{0}, X_{1}, \dots, X_{N-1}]^{T}$. The $N$-point vector is then transformed into the time domain via an inverse discrete Fourier transform (IDFT). A cyclic prefix (CP) of length $N_{\mathrm{CP}}$ is appended to mitigate inter-symbol interference (ISI). The transmitted time-domain sample at index $n$ is given by:
\begin{equation}
s[n] = \frac{1}{\sqrt{N}} \sum_{k=0}^{N-1}  X_k e^{j \frac{2\pi k n}{N}}, \quad -N_{cp} \leq n < N.
\end{equation}

\begin{figure*}[t]
    \centering
    \includegraphics[width=\linewidth]{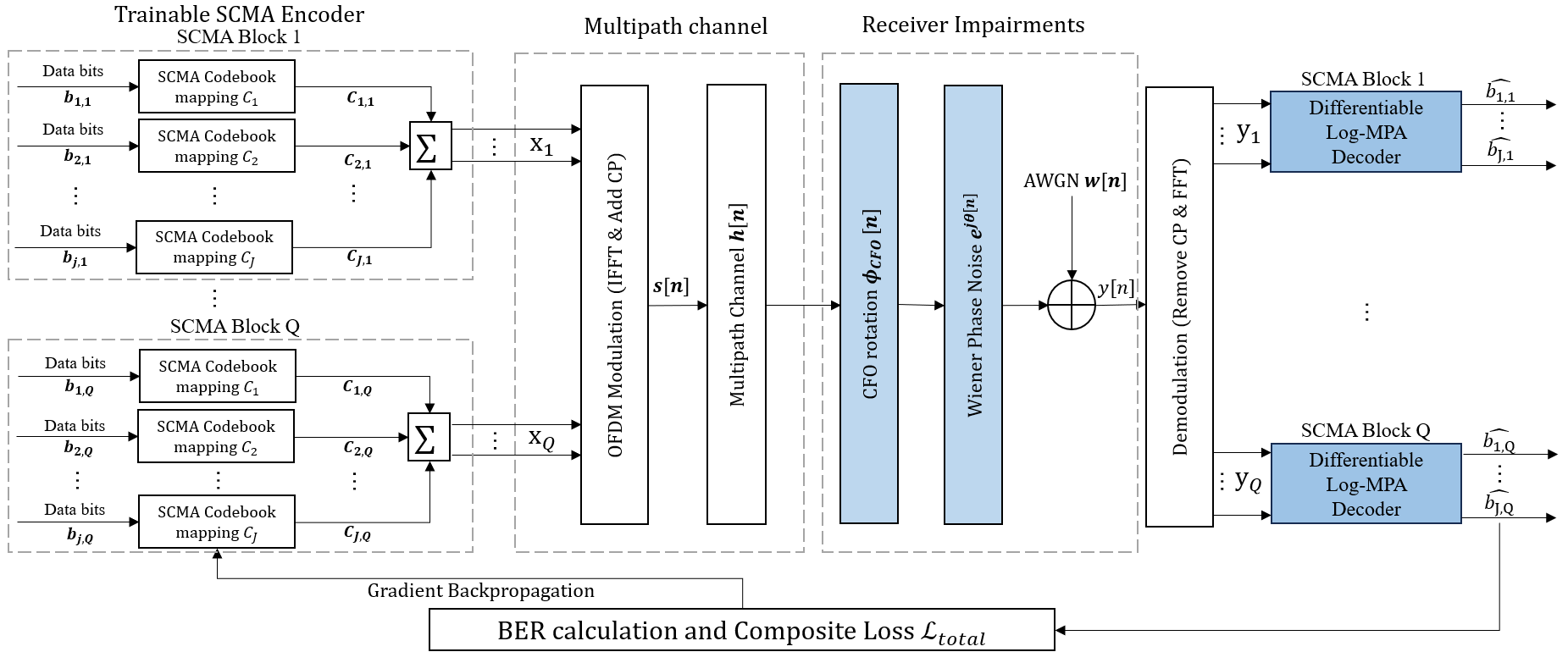}
    \caption{System model of the proposed hardware-aware end-to-end SCMA framework.}
    \label{8-Path Rayleigh Channel wth no CFO and PN}
\end{figure*}

\subsection{Channel Model with Hardware Impairments}
The transmitted signal propagates through a frequency-selective multipath fading channel. Let $h[l]$ denote the discrete-time channel impulse response (CIR) with $L$ taps, where $l \in \{0, \dots, L-1\}$. In practical low-cost transceivers, hardware errors and RF impairments inevitably corrupt the received signal. We explicitly model two critical hardware impairments: CFO and oscillator PN. 

\textbf{CFO:} Due to the mismatch between the local oscillators of the transmitter and the receiver, as well as Doppler shifts, a normalized CFO  $\varepsilon$\footnote{Normalized to the subcarrier spacing.} introduces a time-varying phase rotation:
\begin{equation}
\Phi_{\text{CFO}}[n] = e^{j \frac{2\pi \varepsilon n}{N}}.
\end{equation}

\textbf{PN:} The instability of the local oscillator is modeled as a Wiener process (random walk). The PN $\theta[n]$ at sample $n$ evolves as:
\begin{equation}
\theta[n] = \theta[n-1] + \delta[n], \quad \delta[n] \sim \mathcal{N}(0, \sigma_{\mathrm{PN}}^2),
\end{equation}
where $\delta[n]$ is the real-valued innovation process following a Gaussian distribution with zero mean and variance $\sigma_{\mathrm{PN}}^2$ \cite{A}.

\textbf{Received Signal}: Incorporating the multipath channel, CFO, PN, and additive white Gaussian noise (AWGN), the received time-domain baseband signal $y[n]$ can be expressed as:
\begin{equation}
y[n] = e^{j\theta[n]} \cdot \Phi_{\text{CFO}}[n] \cdot \left( \sum_{l=0}^{L-1} h[l] s[n-l] \right) + w[n],
\label{eq:rx_signal}
\end{equation}
where $w[n] \sim \mathcal{CN}(0, \sigma_{\mathrm{AWGN}}^2)$ represents the AWGN with noise variance $\sigma_{\mathrm{AWGN}}^2$. 

%Note that the proposed model introduces both CFO and Wiener PN at the receiver side. Let $\varepsilon_{tx}$ and $\varepsilon_{rx}$ denote the normalized CFO at the transmitter and receiver, respectively, in reference to the specified carrier frequency in the system. The received signal prior to the receiver CFO is $r[n] = \sum_{l=0}^{L-1} h[l] \big(s[n-l] e^{j 2\pi \varepsilon_{tx} (n-l)/N}\big)$. Rearranging this yields
%\begin{equation}
%\label{eq:rx_signal_expand_cfo}
%r[n] = e^{j \frac{2\pi \varepsilon_{tx} n}{N}} \sum_{l=0}^{L-1} \left( h[l] e^{-j \frac{2\pi \varepsilon_{tx} l}{N}} \right) s[n-l].
%\end{equation}
%Since the multipath channel $h[l]$ is modeled as a zero-mean circularly symmetric complex Gaussian (CSCG) process, multiplying it by the deterministic phase rotation $e^{-j 2\pi \varepsilon_{tx} l/N}$ strictly preserves its statistical distribution. By defining a statistically identical equivalent channel $h'[l] = h[l] e^{-j 2\pi \varepsilon_{tx} l/N}$ and applying the receiver CFO $\varepsilon_{rx}$, the combined signal becomes:
%\begin{equation}
%\label{eq:final_combined_cfo}
%y_{\text{CFO}}[n] = r[n] e^{j \frac{2\pi \varepsilon_{rx} n}{N}} = e^{j \frac{2\pi (\varepsilon_{tx} + \varepsilon_{rx}) n}{N}} \sum_{l=0}^{L-1} h'[l] s[n-l].
%\end{equation}
Note that, although CFO and PN physically arise at both link ends, both can be consolidated at the receiver without loss of fidelity. For CFO, let $\varepsilon_{\mathrm{tx}}$ and $\varepsilon_{\mathrm{rx}}$ be the transmitter and receiver offsets. Since $h[l]$ is a zero-mean circularly symmetric complex Gaussian (CSCG) process, absorbing the transmitter rotation $e^{-j2\pi\varepsilon_{\mathrm{tx}}l/N}$ into a statistically identical equivalent channel leaves only an effective offset $\varepsilon=\varepsilon_{\mathrm{tx}}+\varepsilon_{\mathrm{rx}}$ at the receiver. On the other hand the distributed PN in OFDM downlink is well-approximated by a unified receiver-side model \cite{10746527}. This receiver-side consolidation as shown in Fig.~1 significantly reduces the training complexity while preserving simulation fidelity.

\section{Proposed Hardware-Aware Codebook Design}

In this section, we present a novel hardware-aware end-to-end learning framework. Unlike conventional autoencoders that rely on idealized channel models that assume ideal RF front-ends, we formulate the training process as a stochastic optimization problem that minimizes the expected detection loss averaged over the random realizations of CFO, PN, and noise.

\subsection{Hardware-Aware Computation Graph via Reparameterization}
To enable gradient backpropagation through the stochastic hardware impairments described in Section II, we employ the \textit{reparameterization trick}. This decouples the randomness from the trainable codebook tensor $\boldsymbol{\mathcal{X}} \in \mathbb{C}^{J \times K \times M}$.

Let $\mathbf{z}_{\mathrm{PN}} \sim \mathcal{N}(\mathbf{0}, \mathbf{I}_N)$ and $\mathbf{z}_{w} \sim \mathcal{CN}(\mathbf{0}, \mathbf{I}_N)$ be standard real and complex Gaussian auxiliary variables sampled in each training step, respectively. The stochastic PN process $\boldsymbol{\psi}$ and AWGN $\mathbf{w}$ are reparameterized as differentiable deterministic functions of $\mathbf{z}_{\mathrm{PN}}$ and $\mathbf{z}_{w}$:
\begin{equation}
    \psi_n(\mathbf{z}_{\mathrm{PN}}) = \sum_{i=0}^n \sigma_{\mathrm{PN}} [\mathbf{z}_{pn}]_i, \quad \mathbf{w}(\mathbf{z}_w) = \sigma_{\mathrm{AWGN}} \mathbf{z}_w,
\end{equation}
where $[\mathbf{z}_\mathrm{PN}]_i$ denotes the $i$-th element of vector $\mathbf{z}_\mathrm{PN}$. Consequently, the received time-domain signal $\mathbf{y}$ in the forward pass becomes a differentiable function of the $N$-point frequency-domain vector $\mathbf{X}$:
\begin{equation}
    \mathbf{y}(\mathbf{X}, \varepsilon, \mathbf{z}_{pn}, \mathbf{z}_w) = \mathbf{\Theta}(\boldsymbol{\psi}(\mathbf{z}_{pn})) \mathbf{\Phi}(\varepsilon) \mathbf{H}_{\text{time}} \mathbf{F}^H \mathbf{X} + \mathbf{w}(\mathbf{z}_w),
\end{equation}

where  $\mathbf{F}$ denotes the normalized DFT matrix, and $\mathbf{F}^H$ represents the corresponding IDFT matrix. $\mathbf{H}_{\text{time}}$ is the time-domain multipath channel convolution matrix, and $\mathbf{\Theta}(\cdot)$ and $\mathbf{\Phi}(\cdot)$ are the diagonal phase rotation matrices for PN and CFO, respectively. By sampling $\mathbf{z}$ and $\varepsilon$ in each mini-batch, the autoencoder effectively performs Monte-Carlo integration to approximate the gradient of the expected loss. This explicitly forces the encoder to learn a \textit{rotation-invariant geometric structure} that minimizes the empirical risk averaged over the stochastic phase impairment distributions.

\subsection{Differentiable Mismatched Log-MPA Decoder}
At the receiver, the time-domain signal $\mathbf{y}$ is first stripped of the CP and transformed back to the frequency domain via an $N$-point discrete Fourier transform (DFT), denoted as $\mathbf{Y} = \mathbf{F} \mathbf{y}$. Let $y_k$ be the $k$-th element of $\mathbf{Y}$, corresponding to the received signal at Subcarrier $k$.

In this work, we assume low-cost mMTC sensor receiver which lacks real-time phase tracking and complex ICI equalization. Therefore, it relies on the robust constellation structure learned by the encoder. To maintain no additional online processing beyond standard MPA detection, the receiver employs a \textit{mismatched decoder} that assumes an idealized diagonal frequency response, treating the uncompensated ICI and CPE as equivalent Gaussian noise. We employ an unrolled MPA with $M_{it}$ layers, where $M_{it}$ is the number of iterations. To ensure effective end-to-end differentiability, we utilize the exact log-domain sum-product algorithm (Log-SPA) rather than the max-log approximation. While the max operator is sub-differentiable, the core Log-Sum-Exp (LSE) operator, defined as $\text{LSE}(\mathbf{v}) \triangleq \log(\sum \exp(\mathbf{v}))$, provides a smooth and non-vanishing gradient flow crucial for training deep unrolled networks\footnote{neural networks derived by unfolding iterative algorithms (such as MPA) into trainable layers.}.

Let $x_j \in \mathcal{C}_j$ be the hypothesized symbol for User $j$. Let $h_k$ denote the frequency-domain channel response on Subcarrier $k$. Under the mismatched decoding assumption, since all users share the same multipath channel in the downlink transmission, the message updated from Resource-node $k$ to User-node $j$, evaluated at $x_j$, is formulated as:
\begin{equation}
\begin{split}
    I_{k \to j}(x_j) &= \mathop{\text{LSE}}_{\mathbf{x}_{\mathcal{U}_k \setminus \{j\}}} \Biggl( -\frac{1}{\sigma_{awgn}^2} \left| y_k - h_k \sum_{i \in \mathcal{U}_k} x_i \right|^2 \\
    &\quad + \sum_{i \in \mathcal{U}_k \setminus \{j\}} I_{i \to k}(x_i) \Biggr),
\end{split}
\label{eq:mpa}
\end{equation}
where $\mathbf{x}_{\mathcal{U}_k \setminus \{j\}}$ denotes the set of symbol combinations of all users multiplexed on Subcarrier $k$, excluding User $j$. This differentiable message-passing structure allows gradients to flow seamlessly from the cross-entropy detection loss back to the constellation shaping layer, explicitly optimizing the codebook for the factor graph's behavior under severe uncompensated phase distortions.

\subsection{Theoretically Motivated Composite Loss}
The loss function is designed to optimize the mutual information lower bound while enforcing a curriculum-based performance constraint.

\subsubsection{Task Loss}
The primary task loss $\mathcal{L}_{\text{task}}$ evaluates the empirical risk of uncompensated symbol detection under stochastic hardware impairments. To accelerate convergence and enlarge the decision boundaries of the decoded log-likelihood ratios (LLRs), we formulate $\mathcal{L}_{\text{task}}$ as a weighted sum of the binary cross-entropy ($\mathcal{L}_{\text{BCE}}$) and a soft-margin penalty:
\begin{equation}
    \mathcal{L}_{\text{task}} = \alpha \mathcal{L}_{\text{BCE}} + \beta \mathbb{E} \left[ \ln \left( 1 + \exp \left( \frac{m_0 - (1 - 2b) L_{\mathrm{out}}}{t} \right) \right) \right],
    \label{eq:task_loss}
\end{equation}
The weights are set empirically to $\alpha = 0.3$ and $\beta = 0.8$, with margin parameters $m_0 = 0.8$ and $t = 1.2$. $b \in \{0,1\}$ denoting the target transmitted bit, $L_{\mathrm{out}}$ is the corresponding LLR output by the mismatched Log-MPA, $m_0$ is the decision margin, and $t$ is a 
smoothing parameter that controls the sharpness of the soft-margin penalty, with smaller $t$ yielding a closer approximation to the hard hinge loss. This composite objective forces the autoencoder to evolve rotation-invariant constellations.

\subsubsection{Hinge Loss as a Penalty Method}
We formulate the robustness requirement as a constrained optimization problem: Minimize $\mathcal{L}_{\text{task}}$ subject to the constraint that the performance must surpass a baseline $\mathcal{L}_{\text{base}}$ (e.g., Deka's codebook \cite{deka}) by a margin $m$. Using the exterior penalty method, we transform this inequality constraint into a differentiable penalty term. Let $\Delta \mathcal{L} = \mathcal{L}_{\text{task}} - \mathcal{L}_{\text{base}}$ be the performance gap. The Hinge loss is formulated as:
\begin{equation}
    \mathcal{L}_{\text{hinge}} = \left[ \Delta \mathcal{L} +  m \right]_+^2,
\end{equation}
where $[z]_+ = \max(0, z)$. This term acts as a \textit{curriculum learning} mechanism, guiding the optimization trajectory away from suboptimal local minima associated with conventional designs.
\begin{figure*}[t] 
    \centering

    \begin{subfigure}[b]{0.32\textwidth} 
        \centering
        \includegraphics[width=\linewidth]{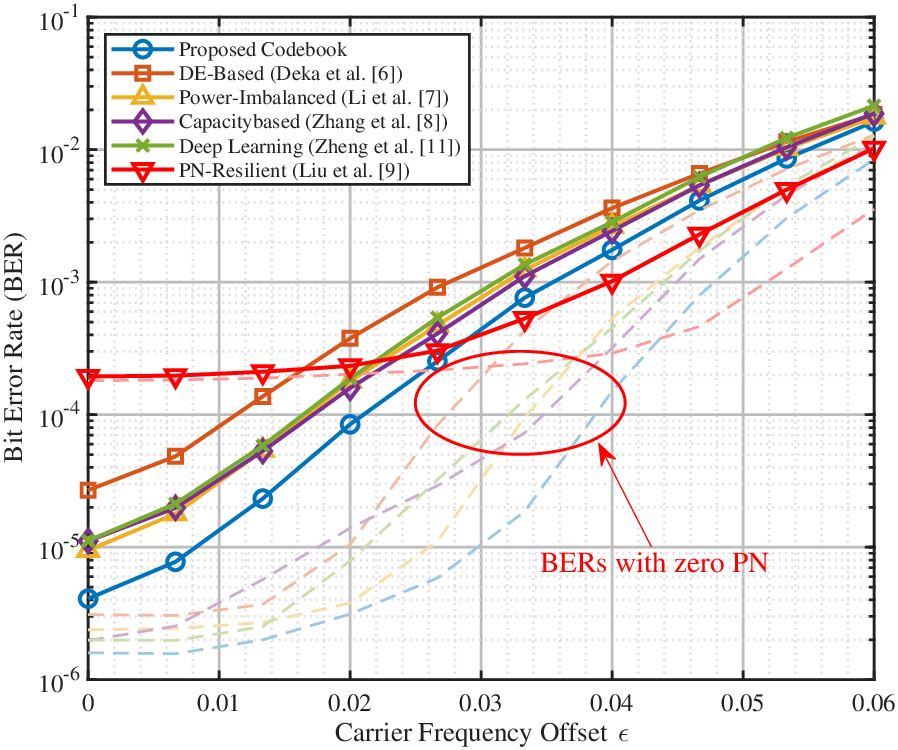} 
        \caption{BER vs. CFO $\varepsilon$ ($\sigma_{pn} = 2.4 \times 10^{-3}$)}
        \label{fig:ber_cfo}
    \end{subfigure}
    \hfil
    \begin{subfigure}[b]{0.32\textwidth}
        \centering
        \includegraphics[width=\linewidth]{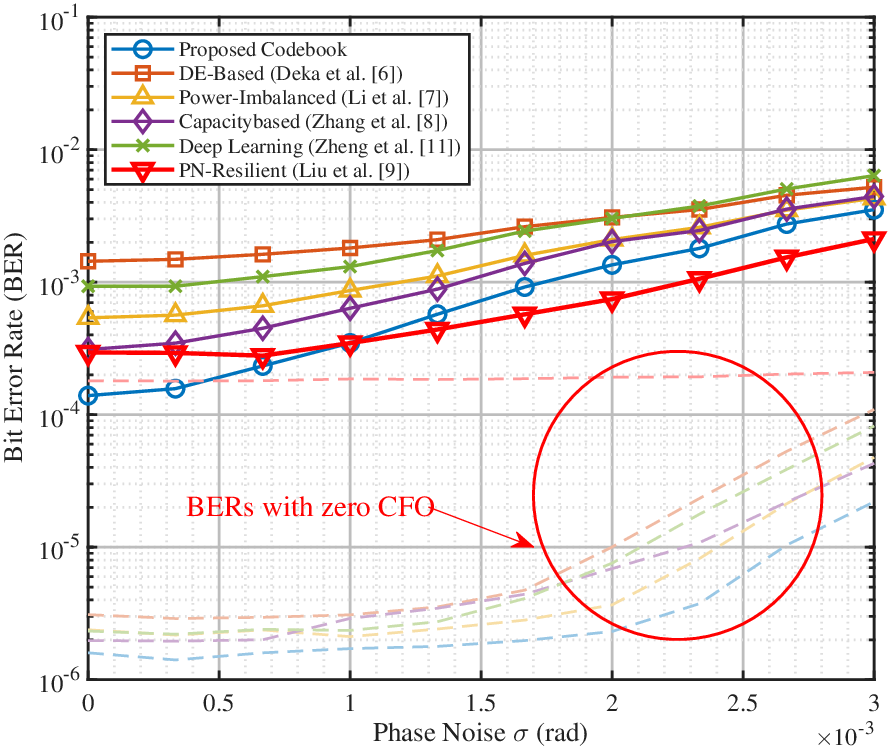} 
        \caption{BER vs. PN $\sigma_{pn}$ ($\varepsilon = 0.04$)}
        \label{fig:ber_pn}
    \end{subfigure}
    \hfill
    \begin{subfigure}[b]{0.32\textwidth}
        \centering
        \includegraphics[width=\linewidth]{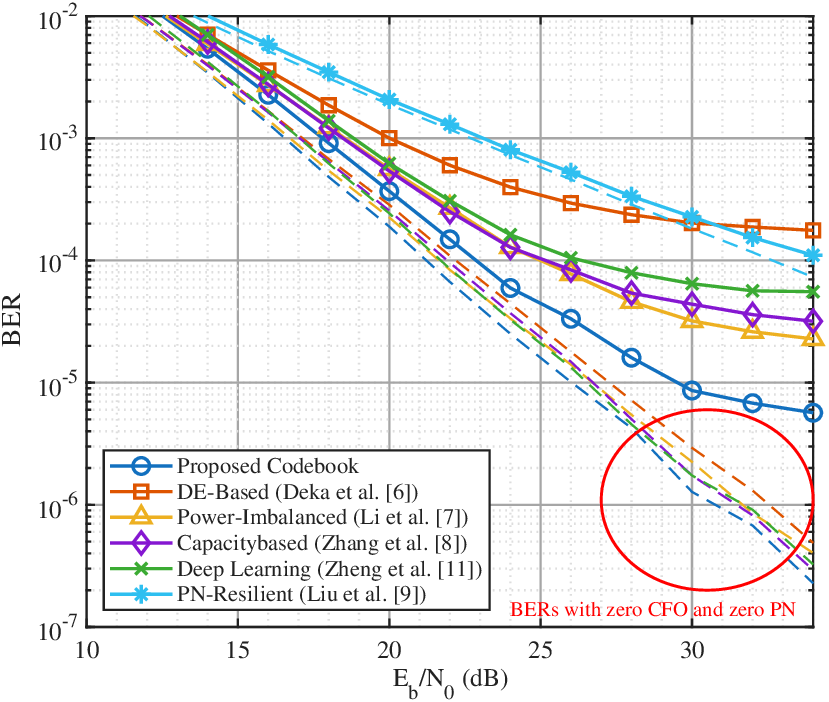} 
        \caption{BER ($\varepsilon=0.03,\sigma_{pn} = 1\times10^{-4}$)}
        \label{fig:ber_snr}
    \end{subfigure}
 \caption{Performance evaluation under hardware impairments. (a) Robustness against CFO, $E_b/N_0$ = 30 dB; (b) Robustness against PN, $E_b/N_0$ = 30 dB; (c) BER versus $E_b/N_0$. Each dashed curve of certain color uses the same codebook as that for the corresponding solid curve.}
    \label{fig:sim_results}
\end{figure*}

\subsubsection{Soft-MED as Union Bound Minimization}
To ensure geometric separability, we maximize the minimum Euclidean distance (MED) of the superimposed constellation. The pairwise error probability (PEP) is upper-bounded by the union bound, which is proportional to $\sum \exp(-d_{i,l}^2/4\sigma_{\mathrm{AWGN}}^2)$. To directly minimize this bound within the gradient descent framework, we define the squared multidimensional Euclidean distance between the $i$-th and $l$-th superimposed codewords as $d_{i,l}^2 = \|\mathbf{x}^{(i)} - \mathbf{x}^{(l)}\|^2$, where $\mathbf{x} \in \mathbb{C}^K$. The Soft-MED loss is formulated over the entire codeword space:
\begin{equation}
    \mathcal{L}_{\text{MED}} = \log \left( \sum_{i \neq l} \exp \left( -\frac{d_{i,l}^2}{\tau} \right) \right).
\end{equation}
Since the Log-Sum-Exp function acts as a smooth upper bound on the maximum function, minimizing $\mathcal{L}_{\text{MED}}$ effectively minimizes the maximum error exponent, thereby maximizing the MED. This ensures the constellation geometry is well-separated prior to channel distortions.

The final objective is $\mathcal{L}_{\text{total}} = \mathcal{L}_{\text{task}} + \lambda_1 \mathcal{L}_{\text{hinge}} + \lambda_2 \mathcal{L}_{\text{MED}}$.

Both $\lambda_1$ and $\lambda_2$ are tuned empirically: $\lambda_1$ is chosen SNR-adaptively in the range $[2, 15]$ so that the hinge penalty is activated whenever $\mathcal{L}_{\text{task}}$ fails to surpass the baseline, while $\lambda_2 \approx 10^{-2}$ keeps the MED term as a soft regularizer that shapes the constellation without dominating the detection loss.

\section{Simulation Results}

In this section, we evaluate the BER performance of the proposed hardware-aware codebooks which are obtained through AE. The simulation setup consists of a downlink SCMA system with $J=6$ users and $K=4$ subcarriers, corresponding to $150\%$ overloading. Further, we assume perfect channel state information (CSI) at the receiver to isolate the impact of CFO and PN. During training, the CFO and PN are fixed at values ($\varepsilon = 0.04$, $\sigma_{\mathrm{PN}} = 1 \times 10^{-3}$). A codebook trained at this single operating point empirically generalizes to milder impairments (Fig. 2(a)-(b)), avoiding the need for impairment randomization during training.

In Figs.~2(a)–(c), the dashed curve of each color is produced by the same codebook as its solid counterpart but with the swept impairment removed. The gap between a solid curve and its dashed reference thus isolates the degradation caused by that impairment, and the proposed codebook stays closest to its own ideal reference across all three scenarios.

\begin{table}[htbp]

\renewcommand{\arraystretch}{1.2}
\caption{Simulation Parameters}
\label{table:parameters}
\centering

\begin{tabular}{@{}ll@{}}
\toprule
\textbf{Parameter} & \textbf{Value} \\ 
\midrule
Channel model               & Frequency-selective Rayleigh fading \\
Number of subcarriers ($N$) & 1024 \\
Cyclic prefix (CP) length   & 32 \\
Power delay profile (PDP)   & 8-tap, exponential decaying \\
Receiver decoder            & Log-MPA \\
Number of iterations ($M_{it}$) & 10 \\

\bottomrule
\end{tabular}
\end{table}

Fig. 2(a) illustrates the BER performance against normalized CFO $\varepsilon$ under severe PN ($\sigma_{\mathrm{PN}} = 2.4 \times 10^{-3}$). While conventional codebooks deteriorate rapidly due to severe ICI, the proposed codebook outperforms at CFO ($\varepsilon < 0.025$). When the CFO value grows larger than 0.025, the PN-resilient
codebook designed in \cite{qu} demonstrates better performance since it was optimized for relatively large CFO. Nevertheless, it suffers from a bit error floor for ideal CFO. In contrast, the proposed codebook offers a balanced performance across the entire evaluated CFO range.
\begin{figure}[t] 
    \centering
    \begin{minipage}{1.0\linewidth} 
        \centering
        \includegraphics[width=\linewidth]{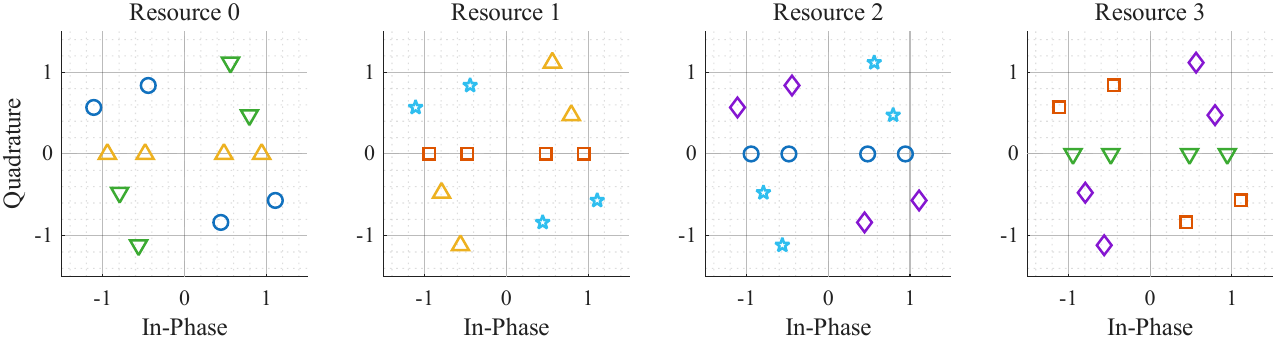}
        \centerline{\footnotesize (a) Deka \textit{et al.} \cite{deka}}
    \end{minipage}
    
    \vspace{0.2cm} 
    
    \begin{minipage}{1.0\linewidth}
        \centering
        \includegraphics[width=\linewidth]{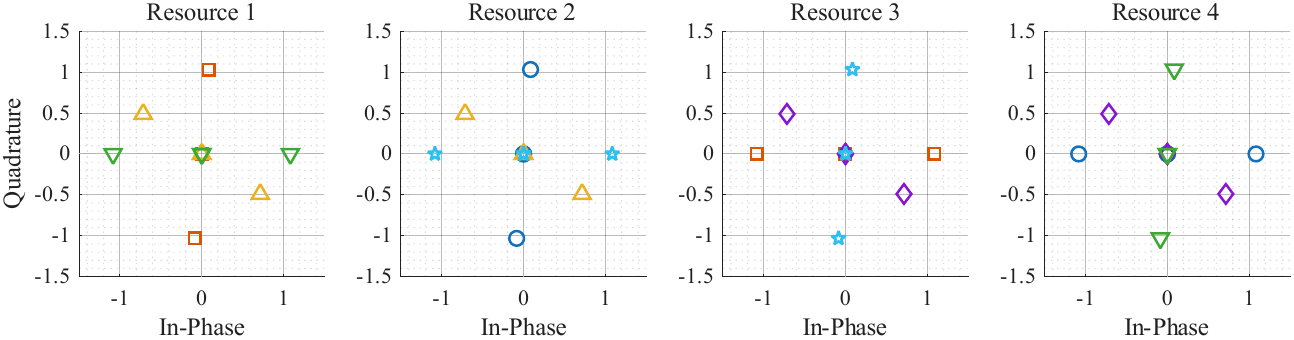}
        \centerline{\footnotesize (b) Liu \textit{et al.} \cite{qu}}
    \end{minipage}
    
    \vspace{0.2cm}
        
    \begin{minipage}{1.0\linewidth}
        \centering
        \includegraphics[width=\linewidth]{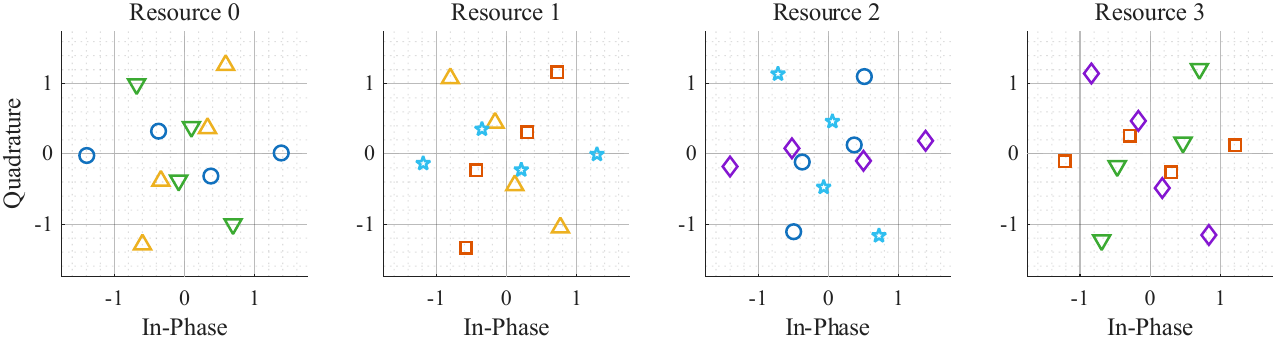}
        \centerline{\footnotesize (c) Proposed hardware-aware codebook}
    \end{minipage}
    
    \caption{Constellation geometries ($J=6$, $K=4$). Each subplot corresponds
    to one resource node; within a subplot, distinct colors and markers denote
    the different users multiplexed on that resource, each contributing $M=4$
    constellation points.}
    \label{fig:constellations_vertical}
\end{figure}

Fig. 2(b) illustrates the BER performance as a function of the PN standard deviation $\sigma_{\mathrm{PN}}$ under a fixed carrier frequency offset of $\epsilon = 0.04$. Across the evaluated range, the proposed codebook consistently outperforms most existing codebooks. Notably, the PN-resilient baseline exhibits a distinct behavior: while it demonstrates superior robustness when $\sigma_{\mathrm{PN}} > 1.3 \times 10^{-3}$, in ideal conditions, the PN-resilient codebook suffers from worse BER compared to other codebooks.

Finally, Fig. 2(c) presents the BER versus $E_b/N_0$ under fixed mixed impairments ($\varepsilon=0.03, \sigma_{\mathrm{PN}}=1\times 10^{-4}$). It is observed that most codebooks suffer from pronounced bit error floors in the high SNR regime ($>25$ dB). This phenomenon arises due to the significant interference induced by CFO and PN. In contrast, the proposed codebook leads to a lower bit error floor, consistently achieving the lowest BER across the entire evaluated SNR range. This validates the effectiveness of the proposed end-to-end training strategy which internalizes the impairment characteristics, optimizing the codebook for enhanced robustness against hardware non-idealities.

To provide physical insights, Fig. 3 visualizes the codebook constellations across four orthogonal resources. As shown in Fig. 3(a), conventional codebooks are optimized for ideal channels but suffer from severe symbol collisions under phase rotations. Conversely, Fig. 3(b) exhibits a distinctly sparse codebook structure. While this sparsity effectively mitigates mutual interference---yielding improved robustness under impairments---it causes significant performance degradation in general conditions, particularly over highly frequency-selective channels. 

In contrast, the proposed codebook converges to a robust \textit{ring-based} geometry, as depicted in Fig. 3(c). Recognizing that CFO and PN primarily cause angular distortions, the model learns to distribute symbols across distinct concentric rings. By maximizing radial separability, it creates a ``phase buffer.'' This ensures that even under dynamic stochastic rotations, the codewords remain sufficiently distinguishable, allowing the proposed scheme to achieve a superior balance between baseline accuracy and extreme-condition resilience. For reference and to facilitate reproducibility, the exact numerical entries of the proposed hardware-aware codebook are explicitly detailed in github.

\section{Conclusion}
This paper proposed a hardware-aware deep learning framework for robust SCMA codebook design for OFDM systems impaired by CFO and PN. By internalizing hardware impairments, the AE-aided SCMA codebook evolves a ring-based geometry that exploits radial separability. Guided by our  union-bound-motivated objective, the proposed design effectively suppresses the error floors observed in conventional baselines. Compared to the standard log-MPA decoder, the proposed codebook  shows no additional complexity, and it offers a highly viable solution for low-cost mMTC with relaxed hardware constraints.

\appendices
\section*{Acknowledgement}
The authors would like to thank Zhiyuan Ma from the University of Essex for his assistance with OFDM simulation code in early stage of this work.
\bibliographystyle{IEEEtran}
\bibliography{refs}
\end{document}